\documentclass[namedreferences]{kluwer}
\usepackage{epsf}

\begin{document}
\begin{article}
\begin{opening}

\title{Comparison of Star Formation Rate estimations \\
from $H_{\alpha}$, FIR and radio data}
\author{Dmitriy \surname{Bizyaev}\email{dmbiz@sai.msu.su}}
\institute{Sternberg Astronomical Institute, Moscow}

\begin{abstract}
We used three indicators of massive star formation:
$H_{\alpha}$, FIR and non-thermal radio luminosities, to compare
estimations of Star Formation Rate (SFR) for the sample of 34 spiral
galaxies.  To adjust SFR values, obtained from different indicators,
we considered the slope $\alpha$
and/or upper mass limit $M_{up}$ of Initial Mass Function (IMF) as
free parameters.  The best agreement between these indicators reaches
for $M_{up} \approx 60 - 100 ~M_{\odot}$ and $\alpha \approx
-3.1 $ in the high mass end of IMF ($M > 10 ~M_{\odot}$). Parallel
with SFR we also estimated FIR excess $X_{FIR}$ defined as a fraction
of the observed FIR which is not related to young massive stars
directly. $X_{FIR}$ is found to be well correlated with types of
spiral galaxies and their colors (B-V): the redder a galaxy the
higher is its FIR excess. We conclude that for any parameters of IMF
the observed FIR flux of early type spiral galaxies needs the
additional source of energy but the massive star radiation.
\end{abstract}

\keywords{galaxies, star formation rate, spiral galaxies}
\end{opening}

\bigskip

One of the most important quantitative  of ongoing star
formation is star formation rate (SFR) defined as a total mass
of stars in a time unit. Young stars reveal themselves in different
spectral bands which allows us to determine SFR using luminosity of
a galaxy in selected spectral bands or lines. One can evaluate SFR
using luminosity of galaxies in emission line, far infrared
continuum (FIR) and in radiocontinuum based on the model results
(see review by Condon, 1992).
These indicators are sensitive to the
presence of massive stars and hence they bear information about
current SFR on scales less than $10^8$ yrs.

Comparison of SFR values obtained from
different indicators enables to constrain some model
parameters, such as present day initial stellar mass function (IMF)
and to make some conclusions about the applicability of the SFR
indicators in galaxies of different types.

Ultraviolet radiation from O and B stars is the source of luminosity
of galaxies $L_{Ha}$ in emission line $H_a$ . Following
Kennicutt (1983), and Dopita, Ryder (1994), we use the equation
\begin{equation}
\label{a1}
L_{H_a} = SFR \int\limits_{m_l}^{M_{up}}l_{H_a}(m)\tau(m)\psi(m)dm ~/~
\int\limits_{m_l}^{M_{up}}m\psi(m)dm
\end{equation}

\noindent Here $\psi\,(m)$ is IMF,
$l_{H_a}(m)$ and $\tau(m)$ are $H_a$ luminosity and
lifetime of stars of mass $m$. To estimate these values the
analytical approximation from Dopita, Ryder (1994) is used:

This approach is sensitive to the extinction of $H_a$ inside of
galaxies. According to Israel, Kennicutt (1980), Kennucutt (1983),
Devereux, Scowen (1994), Devereux et al. (1997), we assume extinction
$A_{Ha} \approx 1^m$.

Significant fraction of stellar bolometric luminosity is re-emitted by
the dust in FIR continuum at 10 -- 1000 $\mu$. FIR flux
is the most sensitive to the presence of stars with mass of order of
several $M_{\odot}$. According to Thronson, Telesco (1986)
\begin{equation}
\label{a2}
L_{FIR} = \int\limits_{m_l}^{M_{up}}P l(m) t_{IR} \psi(m)dm ~/~
\int\limits_{m_l}^{M_{up}} m\psi(m)dm
\end{equation}

\noindent where $P~ \approx ~0.65$ - is a fraction of ionizing
radiation absorbed by the dust, $t_{IR}$ is a time
which a star spends in a dusty star forming region. We assume it to
be $t_{IR} ~=~
10^7$ yrs or $\tau(m)$ if the lifetime is
shorter.  l(m) is bolometric luminosity for stars of main sequence
taken from Mas-Hesse, Kunth (1991).

Synchrotron radio flux may be considered as a third indicator of
star formation in galaxies. One can separate non-thermal
and thermal components due to different slopes of their spectra.
A thermal component at the frequency $\nu$ contributes a fraction of
total luminosity $L_T/L_{total} ~=~ \frac{1.}{1. + 10 ~\nu^(0.1 -
\alpha)}$, where $\alpha ~\approx~ 0.8$ is the slope of synchrotron
spectra (Condon, 1992).

Following Condon (1992) we use following equation which connects
non-thermal luminosity of galaxy
$L_{NT}$ at frequency range
$\nu ~=~$ 0.408 -- 5 GHz with SFR based on calibration from SN Ib, II
rate for our Galaxy (Condon, Yin, 1990, Xu et al., 1994).
\begin{equation}
\label{a3}
L_{NT} = SFR \cdot 1.3 \cdot 10^{23} \nu^{-\alpha}
\int\limits_{M_{SN}}^{M_{up}} \psi(m) dm ~/~
\int\limits_{m_l}^{M_{up}}m \psi(m) dm
\end{equation}
Hereafter we assume $M_{SN} ~=~ 8~ M_{\odot}$.

\section{Comparison of $L_{FIR}/L_{Ha}$ and $L_{NT}/L_{Ha}$ ratios}
Coefficients in formulae (\ref{a1} - \ref{a3}) depend
on upper mass limit
$M_{up}$ and on the slope of IMF in the region of high masses.

Variation of $M_{up}$ allows to explain a wide range of observational
relations $L_{FIR}/L_{Ha}$ in spiral galaxies of different types
(Zasov, 1995).
The slopes of IMF differ from Salpiter's value -2.35 also have been
discussed widely (see reviews in Kennicutt, 1998, Elmegreen, 1999).

We can estimate three independent parameters from equations
(\ref{a1} -- \ref{a3}) by comparing the fluxes taken from
observations in FIR, radio continuum and $H_a$. The first
parameter is SFR which depends also on distances of galaxies.

Another two parameters we evaluate are $X_{FIR}$, the fraction
of cirrus component in general FIR luminosity
which is not connected with current star formation, and one of
parameters describing IMF (either the slope
of IMF in the region of high mass or upper mass limit).
The equations for the observed ratios
$L_{FIR}^{obs}/L_{Ha}^{obs}$ and
$L_{NT}^{obs}/L_{Ha}^{obs}$
may be written as:
\begin{equation}
\label{a4}
(L_{NT}/L_{Ha})^{obs} =
1.3 \cdot 10^{23} \nu^{-\alpha}
\frac{k_{\nu} \int\limits_{M_{SN}}^{M_{up}}\psi(m) dm}
{10^{-0.4A_{H_a}}~ \int\limits_{m_l}^{m_u}l_{H_a}(m)\tau(m)\psi(m)dm}
\end{equation}
\begin{equation}
\label{a5}
(L_{FIR}/L_{Ha})^{obs} =
\frac{(1 ~+~ X_{FIR})~ \int\limits_{m_l}^{M_{up}}P~l(m)~t_{IR}(m)\psi(m)dm}
{10^{-0.4A_{H_a}}~ \int\limits_{m_l}^{M_{up}}l_{H\alpha}(m)\tau(m)\psi(m)dm}
\end{equation}
Here $A_{H_a}$ is the extinction in $H_a$ range,
$k_{\nu} ~=~ 1 ~-~ \frac{1}{1 + 10 \nu^{(0.1 - \alpha)}}$ -
coefficient which enables to extract non-thermal
component of radioflux.

We assume a power law for IMF, following Miller and Scalo,
and Kroupa et al.: $\psi(m) ~=~ m^{-a}$, where
$a = -1.4, ~if~ m ~<~ 1 M_{\odot}$, $a = -2.5, ~if~ 1 ~<~ m ~<~
10 M_{\odot}$ and $a = a_3, ~if~ m ~>~ 10 M_{\odot}$.

Here $a_3$ is taken to be -2.5 if we vary the value of $M_{up}$,
or is variable parameter if $M_{up}$ is taken as a constant.
In the last case we adopt $M_{up} ~=~ 100 M_{\odot}$. The cases
$M_{up} ~=~ 60$ and $120 ~M_{\odot}$ for variable $a_3$ and $a_3 ~=~
-3.1$ for variable $M_{up}$ were also considered.
They give qualitatively similar results.

\section{The sample of galaxies.}
Our sample is based on the list of galaxies observed
at 4.85 and 1.49 GHz and by IRAS (Condon et al., 1995). We excluded the
galaxies which are suspected to have an active nuclear source
("monsters").

\begin{figure}
\epsfxsize=13cm
\epsfbox{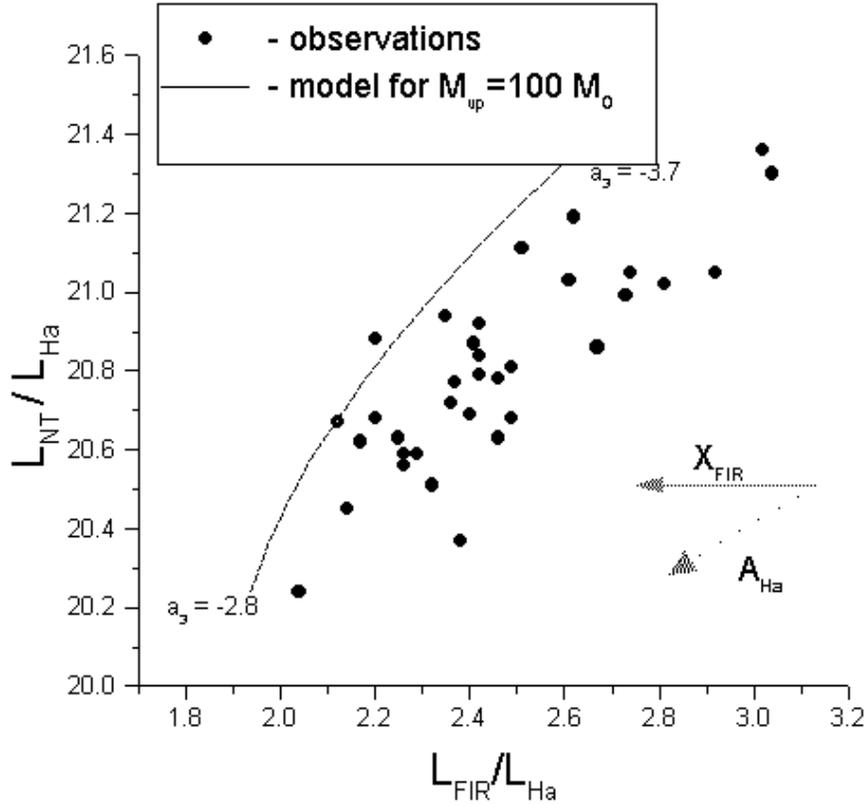}
\caption{Dependence between the observational values of
$L_{FIR} / L_{Ha}$ and $L_{NT} / L_{Ha}$. A solid
line shows the model relationship between $L_{FIR}/L_{Ha}$
and $L_{NT}/L_{Ha}$. High-mass slope of IMF $a_3$ varies
along the line. Arrows show how the corrections for the extinction in
$H_a$ and for the presence of cirrus component in FIR shift the
observational points in the diagram.}
\end{figure}

The observed FIR luminosity in the waverange  1 -- 1000 $\mu$ was
found as $L_{FIR} = 3.65 \cdot 10^5 C(T) D^2
(2.58 S_{60} + S_{100})$, where $S_{60}$ and $S_{100}$ are fluxes at
60 and 100 $\mu$ in Jy.  $L_{FIR}$ is in $L_{\odot}$. $C(T) ~\approx~
1.6$ is a coefficient which depends on color temperature
$S_{60}/S_{100}$. In the observed range of $S_{60}/S_{100}$ C(T)
doesn't change strongly from one galaxy to another (Lonsdale et al.
1995).

The observed $H_a$ luminosities were taken from the data presented by
Kennicutt (1983), Kennicutt et al. (1987), Young et al. (1989),
Romanishin (1990). They were corrected for the presence of [NII]
lines, using relation $L_{Ha + [NII]}/L_{Ha} ~\approx~ 1.5$
(Kennicutt, Kent, 1983).

To minimize selection effects we excluded galaxies with optical
RC3 diameters $D_{25} > 15 '$ from our sample. Also we omitted
galaxies which have galactic latitudes less than $20^o$
to exclude effects of absorption in the Galaxy and
galaxies with the absolute magnitude exceeding $-19^m$).

Finally we have excluded two pairs of galaxies which have
separation between the components more than aperture for $Ha$
observations. The final sample contains 34 galaxies.

\section{Discussion}
Fig.1 shows the dependence between the observational values of
$L_{FIR}^{obs}/L_{Ha}^{obs}$ and $L_{NT}^{obs}/L_{Ha}^{obs}$
(points). The solid line shows the model relationship between
$L_{FIR}/L_{Ha}$ and $L_{NT}/L_{Ha}$.
Parameter $a_3$ varies along the line. Arrows show  how the
corrections for the extinction of $H_a$ and for the presence of
cirrus component of FIR shift the observational points in the
diagram.

\begin{figure}
\epsfxsize=14cm
\epsfbox{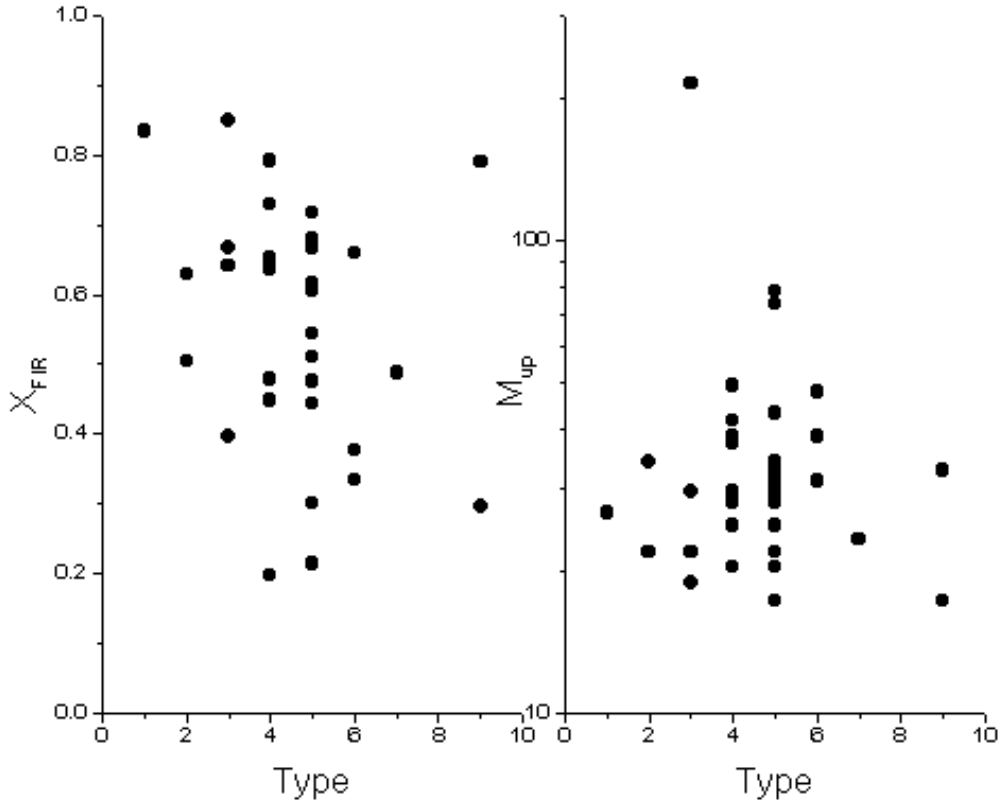}
\caption{Calculated values of $M_{up}$ and $X_{FIR}$ are
plotted against morphological types of galaxies for the case
of variable $M_{up}$.}
\end{figure}

Fig.2 is drawn for the case of variable $M_{up}$. The calculated
values of $M_{up}$ and $X_{FIR}$ are plotted against morphological
types of galaxies of our sample.  $X_{FIR}$ appears to be strongly
dependent on the type and varies from about 0.2 to 0.9. Values
of $M_{up}$ lie between 17 and 78 $M_{\odot}$ (except of NGC 3351
where $M_{up}$ exceeds $200 ~ M_{\odot}$) with the average $M_{up} =
38 ~ M_{\odot}$ and a high dispersion.  Assuming $A_{Ha} ~=~
1^m.5$ instead of $1^m$ we obtain the average $M_{up} = 54
~M_{\odot}$.  In general, $M_{up}$ is dispersed
and is systematically lower than usually accepted values 60 -- 120
$M_{\odot}$.

\begin{figure}
\epsfxsize=14cm
\epsfbox{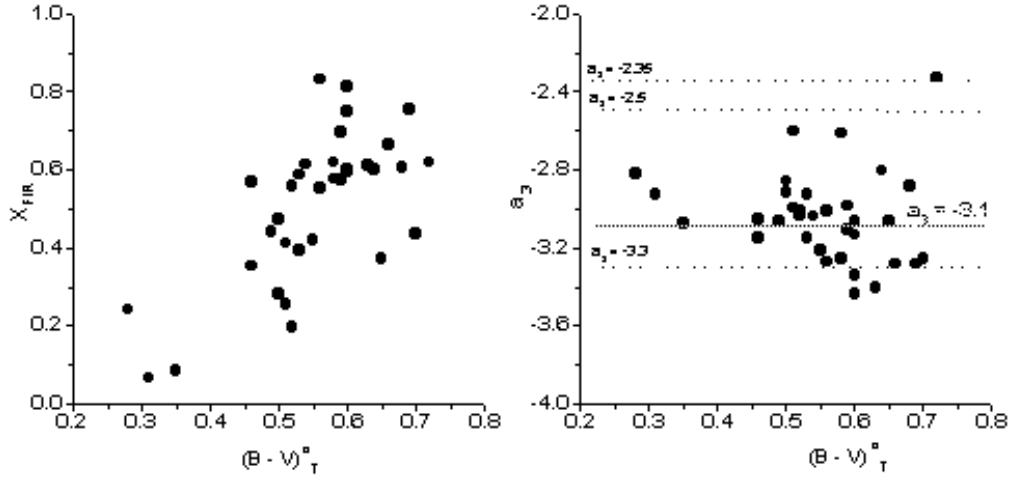}
\caption{Dependences of $X_{FIR}$ and $a_3$
on color indexes $(B-V)^o_T$
for the case of variable $a_3$.
From $80 \%$ to $100 \%$ of FIR radiation traces current star
formation in "blue" galaxies. High-mass slope of IMF $a_3$ seems
to be independent from the integrated colors of the galaxies.}
\end{figure}

A more definite results were found for variable high-mass slope of IMF
$a_3$. From Fig.3 one can see the dependence of $X_{FIR}$ (unlike
$a_3$) on color indexes of galaxies
$(B-V)^o_T$ (from RC3 catalog). For $M_{up} = 100 M_{\odot}$ the
average value $<a_3>=-3.06$, with dispersion 0.20. For $M_{up} = 60
M_{\odot}$ its value is lower ($<a_3>=-2.90$).

The diagram shows that for "blue" galaxies of our sample between
$80 \%$ and $100 \%$ of FIR radiation traces current star
formation. For the galaxies with higher values of
$(B-V)^o_T$ we found that a substantial part
of FIR radiation does not related to star formation.

Hence for "red" early type galaxies with lower SFR  we have
to take into account the part of non-SF cirrus component in FIR
luminosity. Only for "blue" galaxies with $(B-V)^o_T \le 0.5$  with
active star formation $L_{FIR}$ is a good indicator of the presence
of massive stars.

We didn't find a systematical dependencies of $a_3$
and $X_{FIR}$ on luminosities and distances of galaxies, measured
from $V_{GSR}$ velocity (from RC3). It demonstrates the absence of
strong selection effects in the comparison of the observed parameters.
Also there is no correlation between $a_3$, $X_{FIR}$
values and abundance of heavy elements which is available for some
galaxies of our sample.

\begin{acknowledgements}
I would like to thank professor Anotoliy Zasov for initiation
of this work and fruitful discussions and Alexey Moiseev
for discussions and help with the sample. This work was 
partically supported by russian grant RFBR-98-02-17102 and
by grant of Federal program ''Astronomy''.
\end{acknowledgements}

\end{article}
\end{document}